\documentclass[sigconf]{acmart}

\AtBeginDocument{%
  \providecommand\BibTeX{{%
    \normalfont B\kern-0.5em{\scshape i\kern-0.25em b}\kern-0.8em\TeX}}}

\usepackage{CJKutf8}
\setcopyright{acmcopyright}
\copyrightyear{2025}
\acmYear{2025}
\setcopyright{cc}
\setcctype{by}
\acmConference[DIS '25 Companion]{Designing Interactive Systems Conference}{July 5--9, 2025}{Funchal, Portugal}
\acmBooktitle{Designing Interactive Systems Conference (DIS '25 Companion), July 5--9, 2025, Funchal, Portugal}\acmDOI{10.1145/3715668.3736348}
\acmISBN{979-8-4007-1486-3/2025/07}

\begin{document}
\begin{CJK*}{UTF8}{mj}

\title{Beyond Individual UX: Defining Group Experience(GX) as a New Paradigm for Group-centered AI}

\author{Soohwan Lee}
\orcid{0000-0001-8652-3408}
\affiliation{\institution{Department of Design \\ UNIST}
\city{Ulsan}
\country{Republic of Korea}}
\email{soohwanlee@unist.ac.kr}

\author{Seoyeong Hwang}
\orcid{0009-0004-1045-1419}
\affiliation{\institution{Department of Design \\ UNIST}
\city{Ulsan}
\country{Republic of Korea}}
\email{hseoyeong@unist.ac.kr}

\author{Kyungho Lee}
\orcid{0000-0002-1292-3422}
\affiliation{\institution{Department of Design \\ UNIST}
\city{Ulsan}
\country{Republic of Korea}}
\email{kyungho@unist.ac.kr}

\begin{abstract}
Recent advancements in HCI and AI have predominantly centered on individual user experiences, often neglecting the emergent dynamics of group interactions. This provocation introduces Group Experience(GX) to capture the collective perceptual, emotional, and cognitive dimensions that arise when individuals interact in cohesive groups. We challenge the conventional Human-centered AI paradigm and propose Group-centered AI(GCAI) as a framework that actively mediates group dynamics, amplifies diverse voices, and fosters ethical collective decision-making. Drawing on social psychology, organizational behavior, and group dynamics, we outline a group-centered design approach that balances individual autonomy with collective interests while developing novel evaluative metrics. Our analysis emphasizes rethinking traditional methodologies that focus solely on individual outcomes and advocates for innovative strategies to capture group collaboration. We call on researchers to bridge the gap between micro-level experiences and macro-level impacts, ultimately enriching and transforming collaborative human interactions.
\end{abstract}

\ccsdesc[500]{Human-centered computing~HCI theory, concepts and models}
\ccsdesc[300]{Human-centered computing~Collaborative and social computing theory, concepts and paradigms}
\ccsdesc[500]{Human-centered computing~Interaction design theory, concepts and paradigms}

\keywords{human-centered AI; group experience; group-centered AI; social psychology; group dynamics}

\begin{teaserfigure}
  \centering
  \includegraphics[width=1.0\textwidth]{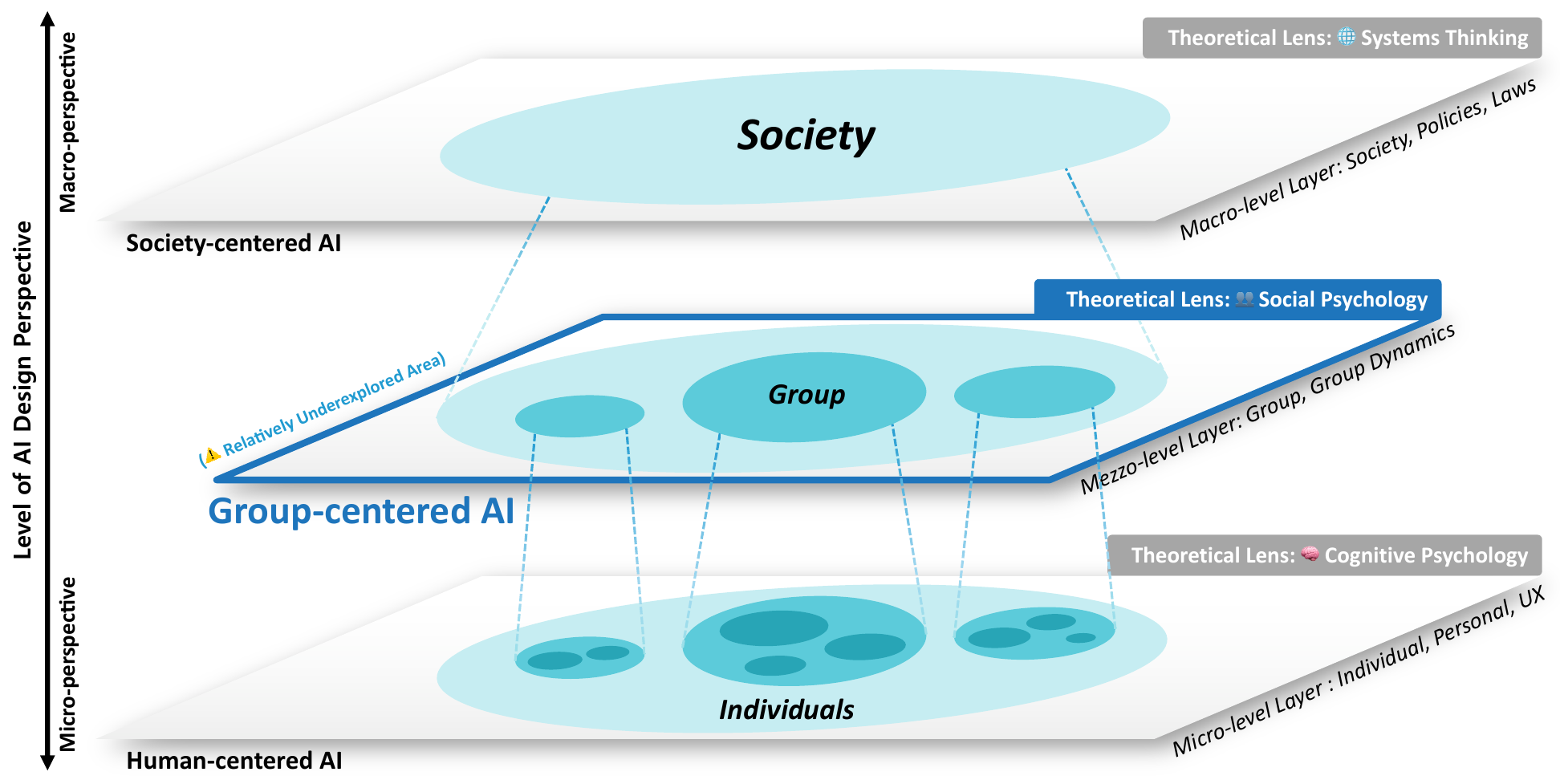}
  \caption{Group-centered AI: A three-tiered framework illustrating the relationship between human-centered AI (micro-level: individual, personal, UX), group-centered AI (mezzo-level: groups, organizations, group dynamics), and society-centered AI (macro-level: society, policies, laws). Dotted lines show connections between entities across levels, demonstrating how group dynamics bridge individual cognition and broader societal structures.}
  \Description{This diagram presents a three-tiered AI design framework spanning micro-, mezzo-, and macro-perspectives: Human-centered AI focuses on individuals and is informed by cognitive psychology; Group-centered AI emphasizes group dynamics and draws from social psychology; Society-centered AI considers societal systems through the lens of systems thinking. Each level—individual, group, and society—is visually layered and interconnected by dotted lines, highlighting how group-level dynamics mediate between personal cognition and broader social structures. The group-centered layer is marked as a relatively underexplored area in current AI research.}
  \label{fig:teaser}
\end{teaserfigure}

\maketitle

\section{Introduction}
The prevalent paradigm in Human-Computer Interaction (HCI), design, and Artificial Intelligence (AI) research has traditionally focused on interactions between technology and individual users. Grounded firmly in cognitive psychology, foundational concepts such as User Experience (UX), usability heuristics, and affordances as prominently advocated by scholars like Donald Norman have fundamentally shaped human-centered design \cite{normanDesignEverydayThings2013, normanCognitiveEngineering1986}. Norman's UX concept specifically underscores the importance of understanding individual users' perceptual, cognitive, and emotional responses to design intuitive, effective, and satisfying technological interactions. This individual-centric perspective has significantly informed the development of Human-Centered AI (HCAI), an approach aiming to create adaptive and personalized AI systems tailored precisely to individual user preferences, contexts, and behaviors. For instance, personalized recommendation systems, virtual assistants, and user-specific interfaces exemplify successes of this individual-oriented paradigm, optimizing efficiency, engagement, and satisfaction by carefully attending to singular user needs and expectations.

However, human-centered AI usually adopts a cognitive perspective, frequently addressing problem situations through a lens that emphasizes interactions between individual users and AI systems \cite{capelWhatHumanCenteredHumanCentered2023a,laiScienceHumanAIDecision2023}. While this cognitive framing excels at capturing nuanced personal experiences, it often overlooks dynamic social interactions and emergent phenomena that uniquely occur within groups \cite{leeExpandingDesignSpace2024}. Even when individual-oriented metrics such as user satisfaction, trust, and task performance are collected from multiple users and aggregated, they do not effectively capture the emergent group-level phenomena arising from collective interaction dynamics. For instance, consider a team using an AI-powered collaborative writing tool: individual satisfaction ratings might suggest overall success, yet these metrics could obscure critical group-level issues such as reinforced power dynamics \cite{houPowerHumanRobotInteraction2024}, marginalization of junior members' contributions \cite{leeConversationalAgentsCatalysts2025a}, or homogenization of the group's creative outputs, phenomena existing exclusively at the collective level \cite{heAIFutureCollaborative2024,chiangAreTwoHeads2023}. In contrast, the emerging paradigm of society-centered AI embraces a systems thinking approach, recognizing the broader socio-technical complexities of AI integration at the societal level. Yet, this macro-level perspective tends to neglect finer-grained interactions, relationships, and processes within smaller groups or collaborative teams. Consequently, current approaches to AI research and design are positioned at two extremes, either excessively micro-focused (individual-oriented human-centered AI) or overly macro-oriented (society-centered AI), and thus fail to address the important intermediate domain of group-level interactions. Although concepts such as Human-AI Teaming have begun to explore collaborative interactions between humans and AI, these frameworks still lack a comprehensive consideration of group decision-making processes, social relationships, and collective experiences that emerge specifically within group contexts. A more integrated and nuanced approach is necessary to bridge this gap, explicitly addressing group-level phenomena in Human-AI interactions.

Therefore, we propose the construct of Group Experience (GX), defined as collective perceptual, emotional, and cognitive responses that emerge specifically when multiple individuals interact cohesively, either with one another or with technology, to accomplish shared tasks. Such limitations of current individualistic paradigms become particularly evident when considering AI systems designed for inherently collective activities such as collaborative decision-making tasks, negotiation and conflict resolution, or creative collaborations. In these contexts, the value emerges not merely from individual optimization, but crucially from how effectively the AI system facilitates collective processes and outcomes. Traditional UX approaches lack the conceptual tools to capture these emergent properties, necessitating a new analytical and design framework centered explicitly on GX. From this perspective, we argue for a fundamental reconceptualization of how we understand, define, and measure human-AI interaction in group contexts through the lens of GX. We suggest that GX represents a distinct construct capable of capturing emergent phenomena occurring when multiple humans and AI systems interact as an integrated socio-technical system. Furthermore, effectively conceptualizing GX requires a multidimensional framework that draws upon foundational theories from social psychology, including social identity theory, social conversation theory, and ETC, organizational behavior, and group dynamics, while simultaneously developing new theoretical constructs tailored specifically to AI-mediated group interactions \cite{myersEBookSocialPsychology2020,forsythGroupDynamics2018}. By adopting this approach, we can bridge the current gap between micro-level individual experiences and macro-level societal considerations, ultimately guiding the development of genuinely group-centered AI systems that enrich collaborative human experiences and support collective endeavors (\autoref{fig:teaser}).
\section{User Experience(UX)? Group Experience(GX)!}
\subsection{What is the Group Experience(GX)?}
We propose the concept of Group Experience(GX), defined as the collective perceptual, emotional, and interactions that uniquely emerge when multiple individuals interact either among themselves or collaboratively with technology as a cohesive unit to accomplish specific tasks. Distinct from traditional User Experience (UX), GX explicitly acknowledges critical group-level phenomena drawn from social psychology and group dynamics, characterized along several interrelated dimensions. GX encompasses \textit{collective sensemaking}, the collaborative process where groups jointly interpret information and construct shared understanding; \textit{group flow and cohesion}, the emergent state of synchronized engagement, collective momentum, and optimal group performance; \textit{social coordination and interaction}, the mechanisms through which groups effectively coordinate actions, manage roles, resolve conflicts, and support communication; \textit{emergent outcomes and behaviors}, novel solutions arising spontaneously from group interactions that individual actions alone cannot predict; and shared identity and equity, the degree to which group members experience belonging, inclusion, fairness, and shared commitment within technology-mediated interactions. Thus, GX represents not merely a concept for multi-user interfaces but a multidimensional construct that explicitly captures emotional resonance, distributed cognition, coordinated actions, and evolving social relationships within groups. Situated between the micro-level focus of individual UX and macro-level societal impacts, we expect that GX provides essential guidance for designing technologies that authentically enhance collaborative processes and collective human interactions.

\subsection{Cases: Good for UX, Bad for GX?}
Individual-centered UX principles grounded in cognitive psychology can effectively enhance personal efficiency, satisfaction, and task performance. However, optimizing UX for individual users may inadvertently lead to negative impacts at the group level, degrading the overall GX. Several cases illustrate this tension clearly. These examples highlight critical shortcomings of individual-centered UX approaches in collaborative contexts. Despite individual-level benefits, they overlook essential group dynamics, reinforcing the need for a dedicated GX framework to design truly group-centered interactive systems

One representative example is a performance system focused on immediate public personal feedback. Individually, users benefit from rapid and precise feedback, enabling quick personal growth and skill development \cite{trajkovaTakesTutuBallet2018}. However, when such personalized feedback becomes excessively emphasized or is publicly shared, it can inadvertently promote self-centered behaviors, prioritizing individual achievements over group success \cite{leeExpandingDesignSpace2024,doHowShouldAgent2022}. Moreover, negative feedback shared publicly may lead to feelings of shame or embarrassment, undermining psychological safety and harming the overall group climate and collaborative interactions.

Another illustrative case is an anonymous voting and feedback system, designed to ensure psychological safety and encourage honest expression. Individually, users initially feel more comfortable sharing opinions without fear of social repercussions \cite{jessupEffectsAnonymityGDSS1990a}. Yet, extensive anonymity may erode interpersonal accountability and mutual trust, complicating consensus-building and conflict resolution within the group. Overreliance on anonymity may generate suspicion and perceived unfairness among members, obstructing the establishment of cohesive group norms and effective collaboration \cite{leeAmplifyingMinorityVoices2025a}.

\begin{figure*}[]
  \centering
  \includegraphics[width=1.0\textwidth]{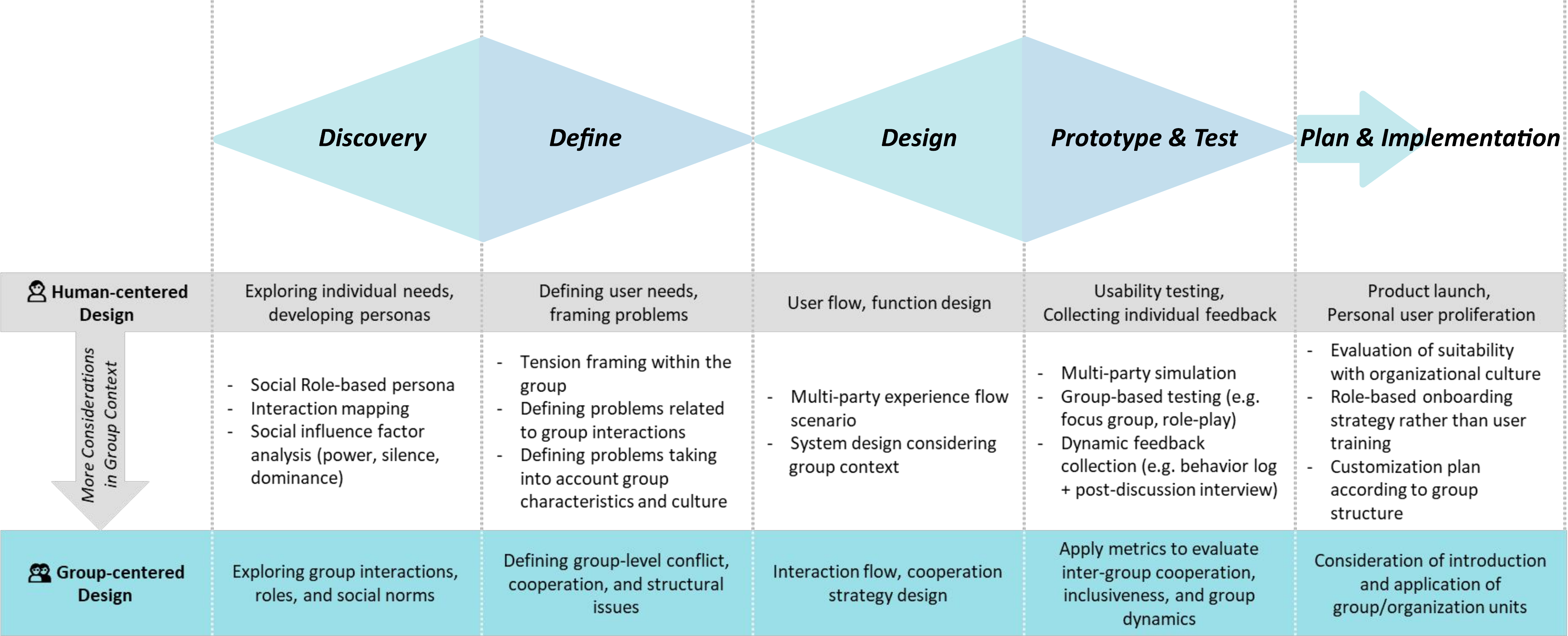}
  \caption{Group-centered Design: Comparison between Human-centered Design and Group-centered Design approaches across five design phases (Discovery, Define, Design, Prototype \& Test, Plan \& Implementation). The diagram illustrates how traditional HCD methods focused on individual needs transition to GCD approaches that explicitly address group dynamics, social roles, collective interactions, and organizational structures throughout the design process.}
  \Description{This diagram compares Human-centered Design (HCD) and Group-centered Design (GCD) across five phases of the design process: Discovery, Define, Design, Prototype \& Test, and Plan \& Implementation. HCD focuses on individual needs—such as personas, user flows, and individual feedback—while GCD expands each phase to address group-level aspects, including social roles, group tensions, collective decision-making, and organizational structures. Each column represents a phase, with GCD considerations layered beneath the traditional HCD approach, highlighting methods like group-based simulations, role-based onboarding, and metrics for evaluating inter-group dynamics and cooperation.}
  \label{fig:groupCenteredDesign}
\end{figure*}

\subsection{Towards Group-centered Design Approach}
Group-centered design (GCD) extends traditional human-centered design (HCD) \cite{jacobsonInformationDesign2000} by explicitly emphasizing the collective interactions, relationships, and contextual dynamics of groups as design considerations (\autoref{fig:groupCenteredDesign}). While HCD inherently includes understanding individual needs, goals, and behaviors within social contexts, it typically addresses these elements from an individual's perspective. However, when interactions involve multiple users who collectively influence each other, individual-level insights alone may fall short of capturing the full complexity of group interactions. GCD thus builds upon HCD by incorporating detailed analysis of group-level phenomena, such as social roles, power dynamics, communication patterns, and shared norms, into the core design process.

In practice, whereas HCD typically employs individual personas and user journeys to represent distinct users, GCD introduces role-based personas and interaction mappings that reflect the varied roles individuals assume within groups, such as opinion leaders, facilitators, or marginalized voices. Methodologies like group interviews, role-based workshops, social network analysis, and conflict mapping enrich the traditional individual-centered techniques, allowing designers to more explicitly address issues such as group cohesion, equity in participation, and psychological safety. Furthermore, prototyping and evaluation within GCD extend conventional HCD approaches by adopting group-level methodologies, such as collective role-play simulations and collaborative scenario testing. Metrics expand beyond individual usability or satisfaction to encompass collective measures like collaboration quality, diversity of expressed ideas, distribution of participation, and legitimacy of group decisions. Therefore, GCD does not replace but rather enhances the traditional HCD approach, emphasizing that genuinely effective design for groups requires deliberate consideration of the collective experiences and interactions that emerge distinctly within group contexts.

\section{Beyond Individuals: A New AI Paradigm for Groups}
\subsection{What is the Group-centered AI?}
Building directly upon the GX framework established earlier, we propose Group-centered AI (GCAI) as the technological manifestation designed to nurture and enhance collective experiences. Where GX identifies the perceptual, emotional, and cognitive dimensions that emerge in group contexts, GCAI provides the mechanisms through which AI can actively support these phenomena. This novel approach focuses on the mezzo-level between individual and societal AI paradigms, explicitly supporting, augmenting, and optimizing group interactions by recognizing collectives as meaningful social units beyond the sum of individuals. Unlike Human-centered AI that prioritizes individual autonomy and Society-centered AI that addresses broad societal welfare, GCAI enhances specific group processes, including collaboration, decision-making, consensus-building, and conflict resolution (\autoref{fig:teaser}). Drawing from social psychology, organizational behavior, Computer-Supported Cooperative Work, and participatory design, GCAI comprises five core elements: (1)  Interaction Mediation facilitates constructive conflict resolution and consensus-building by moderating group tensions and guiding interactions toward mutual understanding \cite{claggettRelationalAIFacilitating2025,doanDesignSpaceOnline2025,phamEmbodiedMediationGroup2024,shinChatbotsFacilitatingConsensusBuilding2022,kimBotBunchFacilitating2020}; (2) Social Transparency enhances collective awareness about individual contributions, roles, intentions, and hidden social dynamics, empowering groups to recognize patterns and adjust their interactions accordingly \cite{liaoDesigningResponsibleTrust2022,doHowShouldAgent2022,doInformExplainControl2023}; (3) Adaptive Scaffolding dynamically supports cognitive and social tasks like collaborative problem-solving, negotiation, and consensus formation by adapting interventions to evolving group contexts rather than merely personalizing for individuals \cite{wangSocialRAGRetrievingGroup2025a,chenIntegratingFlowTheory2024b,liuProactiveConversationalAgents2025}; (4) Collective Intelligence Amplification strategically aggregates diverse perspectives, explicitly amplifies minority or critical voices, and encourages constructive dissent through intelligent provocations that prevent premature convergence \cite{leeAmplifyingMinorityVoices2025a, zhangBreakingBarriersBuilding2025a,heAIFutureCollaborative2024,houdeControllingAIAgent2025b,hwangSoundSupportGendered2024,leeConversationalAgentsCatalysts2025a}; and (5) Group-Level Accountability and Ethics ensures that ethical considerations and accountability mechanisms operate at the collective level, fostering group reflection and ethical deliberation beyond individual responsibility. These principles apply across domains like collaborative decision-making, education, remote work, and social robotics, where traditional AI approaches often overlook the unique challenges of group-level interaction.

GCAI demands a fundamental shift in how we design, evaluate, and implement AI systems. Current AI systems like recommendation engines optimize for individual preferences, neglecting group dynamics and collective experiences. GCAI reframes Human-AI teaming beyond dyadic relationships to address complex intra-group dynamics, including majority influence, persuasion, and collective deliberation. This paradigm requires expanding ethical frameworks to explicitly consider group-level concerns such as balancing majority and minority perspectives, preventing dominance reinforcement, and fostering inclusive participation. Methodologically, GCAI necessitates moving from conventional human-centered design toward group-focused approaches, including multi-user interaction scenarios and evaluation metrics that capture collective experiences like group cohesion, psychological safety, and diversity of expressed ideas. By systematically addressing the complexity of group interactions, GCAI creates a pathway for AI systems that genuinely support human collaboration and collective agency, filling a critical gap between individually-focused and society-level AI paradigms. This conceptual reorientation enables us to design intelligent systems that enhance rather than diminish the richness of collective human experience.

\subsection{Provocations for the Future of Group-AI Interaction}
Introducing Group-centered AI (GCAI) opens compelling new directions and provocative considerations for future research and design. To systematically enrich and expand this emerging paradigm, researchers and designers must grapple with critical challenges, including:

\begin{itemize}
    \item \textbf{\textit{Balancing Individual and Collective Interests:}} GCAI must effectively mediate tensions between individual autonomy and group consensus. Ethical guidelines and design principles should clearly address how AI navigates trade-offs when personal preferences conflict with collective goals.
    \item \textbf{\textit{Optimal Levels of Social Transparency and AI Intervention:}} Researchers need to determine the appropriate extent of AI transparency and intervention. Effective GCAI systems should balance active mediation of harmful group dynamics with maintaining group agency and autonomy.
    \item \textbf{\textit{Authority, Agency, and Trust in Adaptive Scaffolding:}} Designers should carefully examine how authoritative or directive AI systems can become in group decision-making contexts. Understanding how such interventions affect group trust, psychological safety, and perceived fairness is essential.
    \item \textbf{\textit{Constructive Provocations and Dissent Management:}} Future GCAI systems must introduce intelligent provocations that productively stimulate critical thinking and diverse viewpoints, without inadvertently escalating conflicts or negatively affecting group harmony.
    \item \textbf{\textit{Ethical Accountability at the Group Level:}} GCAI requires novel accountability mechanisms explicitly designed for collective contexts. Researchers should explore how AI can foster shared ethical reflection, accountability, and responsibility among group members.
    \item \textbf{\textit{Methodological Innovations for Group-Level Evaluation:}} New metrics and methodologies must be developed to rigorously assess collective experiences such as group cohesion, psychological safety, and diversity of participation. Moving beyond traditional individual-level evaluations is critical for fully realizing GCAI.
\end{itemize}

In addressing these provocations, researchers and designers have a profound opportunity to significantly enrich the design space for AI systems. Through systematic engagement with these provocations, the field can create AI systems that meaningfully amplify the richness, complexity, and potential of human group interaction.
\section{Conclusion}
We introduce Group Experience (GX) as a paradigm for collective perceptual, emotional, and interactive dimensions in group interactions. We challenge individual-centric design, advocating for group-centered AI (GCAI) that mediates group processes, amplifies diverse voices, and maintains collective accountability. Key questions include balancing individual autonomy with group cohesion, calibrating AI intervention, and developing metrics for collaboration. We urge integrating insights from social psychology, organizational behavior, and HCI to transform how we support and evaluate collective interactions. We call for innovative strategies and ethical frameworks bridging individual experiences and group phenomena to enrich interactive, group-centered AI systems.

\begin{acks}
This research was partially supported by a grant from the Korea Institute for Advancement of Technology (KIAT) funded by the Government of Korea (MOTIE) (P0025495, Establishment of Infrastructure for Integrated Utilization of Design Industry Data). This work was also partially supported by the Technology Innovation Program (20015056, Commercialization design and development of Intelligent Product-Service System for personalized full silver life cycle care) funded by the Ministry of Trade, Industry \& Energy(MOTIE, Korea).
\end{acks}

\bibliographystyle{ACM-Reference-Format}
\bibliography{devilsAdvocate, devilsAdvocate_workshop, HCAI, group}

\end{CJK*}
\end{document}